\documentclass[aps,prl,twocolumn,showpacs,floatfix]{revtex4}

\usepackage{graphicx}
\usepackage[usenames]{color}

\newcommand{\met}       {\mbox{$\not\!\!E_T$}}
\newcommand{\rar}       {\rightarrow}
\newcommand{\rargap}    {\mbox{ $\rightarrow$ }}
\newcommand{\ppbar}     {\mbox{$p\bar{p}$}}
\newcommand{\ttbar}     {\mbox{$t\bar{t}$}}

\newcommand{\comphep}   {\sc comphep}
\newcommand{\singletop} {\sc singletop}
\newcommand{\pythia}    {\sc pythia}
\newcommand{\alpgen}    {\sc alpgen}

\newcommand{\geant}     {\sc geant}

\begin{document}

\title{Observation of single top-quark production}

%
\author{V.M.~Abazov$^{36}$}
\author{B.~Abbott$^{74}$}
\author{M.~Abolins$^{64}$}
\author{B.S.~Acharya$^{29}$}
\author{M.~Adams$^{50}$}
\author{T.~Adams$^{48}$}
\author{E.~Aguilo$^{6}$}
\author{M.~Ahsan$^{58}$}
\author{G.D.~Alexeev$^{36}$}
\author{G.~Alkhazov$^{40}$}
% ...

\affiliation{\vspace{0.1 in}(The D\O\ Collaboration)\vspace{0.1 in}}
% ...
\affiliation{$^{6}$University of Alberta, Edmonton, Alberta, Canada;
                Simon Fraser University, Burnaby, British Columbia, Canada;
                York University, Toronto, Ontario, Canada and
                McGill University, Montreal, Quebec, Canada}
% ...
\affiliation{$^{29}$Tata Institute of Fundamental Research, Mumbai, India}
% ...
\affiliation{$^{36}$Joint Institute for Nuclear Research, Dubna, Russia}
% ...
\affiliation{$^{40}$Petersburg Nuclear Physics Institute,
                St. Petersburg, Russia}
% ...
\affiliation{$^{48}$Florida State University, Tallahassee, Florida 32306, USA}
% ...
\affiliation{$^{50}$University of Illinois at Chicago,
                Chicago, Illinois 60607, USA}
% ...
\affiliation{$^{58}$Kansas State University, Manhattan, Kansas 66506, USA}
% ...
\affiliation{$^{64}$Michigan State University,
                East Lansing, Michigan 48824, USA}
% ...
\affiliation{$^{74}$University of Oklahoma, Norman, Oklahoma 73019, USA}
% ...

\date{March 4, 2009}

\begin{abstract}

We report observation of the electroweak production of single
top quarks in {\ppbar} collisions at $\sqrt{s} = 1.96$~TeV based on
2.3~fb$^{-1}$ of data collected by the D0 detector at the Fermilab
Tevatron Collider. Using events containing an isolated electron or
muon and missing transverse energy, together with jets originating
from the fragmentation of $b$ quarks, we measure a cross section of
$\sigma({\ppbar}{\rargap}tb+X,~tqb+X) = 3.94 \pm 0.88$~pb. The
probability to measure a cross section at this value or higher in the
absence of signal is $2.5\times10^{-7}$, corresponding to a
5.0~standard deviation significance for the observation.

\end{abstract}


\maketitle 


At hadron colliders, top quarks can be produced in pairs via the
strong interaction or singly via the electroweak
interaction~\cite{singletop-willenbrock}. Top quarks were first
observed via pair production at the Fermilab Tevatron Collider in
1995~\cite{top-obs-1995}. Since then, pair production has been used to
make precise measurements of several top quark properties, including
the top quark mass~\cite{top-mass-properties}. Single top quark
production, on the other hand, serves as a probe of the $Wtb$
interaction~\cite{singletop-wtb}, and its production cross section
provides a direct measurement of the magnitude of the quark mixing
matrix element $V_{tb}$ without assuming three quark
generations~\cite{singletop-vtb-jikia}. However, measuring the yield
of single top quarks is difficult because of the small production rate
and large backgrounds.

In 2007, we presented the first evidence for single top quark
production and the first direct measurement of
$|V_{tb}|$~\cite{d0-prl-2007,d0-prd-2008} using 0.9~fb$^{-1}$ of
Tevatron data at a center-of-mass energy of 1.96~TeV. Recently, the
CDF collaboration has also presented such evidence in 2.2~fb$^{-1}$ of
data~\cite{cdf-prl-2008}. This Letter describes the observation
of a single top quark signal in 2.3~fb$^{-1}$ of Tevatron data. The
CDF collaboration is also reporting observation of single top quark 
production~\cite{Aaltonen:2009jj}.

\begin{figure}[!h!tbp]
\includegraphics[width=3in]{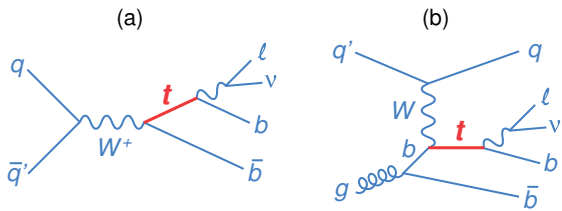}
\vspace{-0.1in}
\caption[feynman]{Representative Feynman diagrams for (a) $s$-channel
single top quark production and (b) $t$-channel production, showing
the top quark decays of interest.}
\label{feynman}
\end{figure}

When top quarks are produced singly, they are accompanied by a bottom
quark in the $s$-channel production mode~\cite{singletop-cortese} or
by both a bottom quark and a light quark in $t$-channel
production~\cite{singletop-willenbrock,singletop-yuan}, as illustrated
in Fig.~\ref{feynman}. We search for both of these processes at once.
The sum of their predicted cross sections is $3.46 \pm
0.18$~pb~\cite{singletop-xsec-kidonakis} for a top quark mass $m_t =
170$~GeV, at which this analysis is performed. We refer to the
$s$-channel process as ``$tb$'' production, where $tb$ includes both
$t\bar{b}$ and $\bar{t}b$ states. The $t$-channel process is
abbreviated as ``$tqb$,'' where this includes $tq\bar{b}$,
$t\bar{q}\bar{b}$, $\bar{t}\bar{q}b$, and $\bar{t}qb$ states.

The analysis presented in this Letter is an improved version of the
one from 2007~\cite{d0-prl-2007,d0-prd-2008}, with a larger
dataset. Most definitions and abbreviations used here are explained in
detail in Ref.~\cite{d0-prd-2008}. The measurement focuses on the
final state containing one high transverse momentum ($p_T$) lepton
($\ell$ = electron or muon) not near a jet (``isolated''), large
missing transverse energy ({\met}) indicative of the passage of a
neutrino $\nu$, a $b$-quark jet from the decay of the top quark
($t{\rar}Wb{\rar}\ell \nu b$), and possibly another $b$~jet and a
light jet as indicated above. The data were collected with the D0
detector~\cite{d0-detector} using a logical OR of many trigger
conditions in place of only the single-lepton plus jets triggers used
previously. Several offline selection criteria, including $b$-jet
identification requirements for double-tagged events, have been
loosened. These improvements have increased the signal acceptance by
18\%. The backgrounds are $W$~bosons produced in association with
jets, top quark pair ({\ttbar}) production with decay into the
lepton+jets and dilepton final states (when a jet or a lepton is not
reconstructed), and multijet production, where a jet is
misreconstructed as an electron or a heavy-flavor quark decays to a
muon that passes isolation criteria. $Z$+jets and diboson processes
form minor additional background components.


We consider events with two, three, or four jets (which allows for
additional jets from initial-state and final-state radiation),
reconstructed using a cone algorithm in $(y,\phi)$ space, where $y$ is
the rapidity and $\phi$ is the azimuthal angle, and the cone radius is
0.5~\cite{d0-prd-2008}. The highest-$p_T$ (leading) jet must have
$p_T>25$~GeV, and subsequent jets have $p_T>15$~GeV; all jets have
pseudorapidity $|\eta|<3.4$. We require $20 < {\met} < 200$~GeV for
events with two jets and $25 < {\met} < 200$~GeV for events with three
or four jets. Events must contain only one isolated electron with
$p_T>15$~GeV and $|\eta|<1.1$ ($p_T>20$~GeV for three- or four-jet
events), or one isolated muon with $p_T>15$~GeV and $|\eta|<2.0$. The
background from multijets events is kept to $\approx$5\% by requiring
high total transverse energy and by demanding that the {\met} is not
along the direction of the lepton or the leading jet. To enhance the
signal fraction, one or two of the jets are required to originate from
long-lived $b$~hadrons. We achieve this goal by using a neural network
(NN) $b$-jet tagging algorithm~\cite{btagging-scanlon}. The variables
used to identify such jets rely on the characteristics of a secondary
vertex and tracks with large impact parameters. After $b$-jet
identification, we require the leading $b$-tagged jet to have
$p_T>20$~GeV. To further improve the sensitivity, we split the data by
lepton flavor, number of jets and $b$-tagged jets, and data collection
period.


We model the signal using the {\comphep}-based next-to-leading order
(NLO) Monte Carlo (MC) event generator
{\singletop}~\cite{singletop-mcgen}. The decays of the top quark and
resulting $W$~boson, both with standard model (SM) widths, are modeled
in {\singletop} to preserve spin information. {\pythia}~\cite{pythia}
is used to model the hadronization of generated partons. We assume the
SM prediction for the ratio of the $tb$ and $tqb$ cross
sections~\cite{singletop-xsec-kidonakis}.


The {\ttbar}, $W$+jets, and $Z$+jets backgrounds are simulated using
the {\alpgen} leading-log MC event generator~\cite{alpgen} and
{\pythia} to model hadronization. The {\ttbar} background is
normalized to the predicted cross section~\cite{ttbar-xsec}. The
diboson backgrounds are modeled using {\pythia}. In the simulation of
the $W$+jets backgrounds, we scale the {\alpgen} cross sections for
events with heavy flavor jets by factors derived from calculations of
NLO effects~\cite{mcfm}: $Wb\bar{b}$ and $Wc\bar{c}$ are scaled by
1.47, and $Wcj$ by 1.38.

All MC events are passed through a {\geant}-based simulation of the D0
detector and are reconstructed using the same software as for the
data. Data events from random beam crossings are overlaid on the
simulation to better model the effects of detector noise and multiple
{\ppbar} interactions. Small differences between data and simulation
in the lepton and jet reconstruction efficiencies and resolutions are
corrected in the simulation as measured from separate data samples.
We also correct the $\eta({\rm jets})$, $\Delta\phi({\rm jet1,jet2})$,
and $\Delta\eta({\rm jet1,jet2})$ distributions in the $W$+jets
samples to match data.

The multijets background is modeled using independent data samples
containing leptons that are not isolated. The multijets background,
combined with the background from $W$+jets, is normalized to the
lepton+jets data with other backgrounds subtracted, using the
$p_T(\ell)$, {\met}, and the $W$~boson transverse mass distributions
before $b$-jet identification is applied.

The $b$-tagging algorithm is modeled in simulated events by applying
weights (``tag-rate functions'') measured from data that account for
the probability for each jet to be tagged as a function of jet flavor,
$p_T$, and $\eta$. After $b$~tagging, an empirical correction of $0.95
\pm 0.13$ for the $Wb\bar{b}$ and $Wc\bar{c}$ fractions is derived
from the $b$-tagged and not-$b$-tagged two-jet data and simulated
samples.


The above selections give 4,519 $b$-tagged lepton+jets events, which
are expected to contain $223\pm 30$ single top quark
events. Table~\ref{event-yields} shows the event yields, separated by
jet multiplicity. The acceptances are $(3.7 \pm 0.5)\%$ for $tb$ and
$(2.5 \pm 0.3)\%$ for $tqb$, expressed as percentages of the inclusive
single top quark production cross section in each channel.

\begin{table}[!h!btp]
\vspace{-0.1in}
\caption[eventyields]{Number of expected and observed events in
2.3~fb$^{-1}$ for $e$ and $\mu$, and 1 and 2 $b$-tagged analysis
channels combined. The uncertainties include both statistical and
systematic components.}
\label{event-yields}
\begin{ruledtabular}
\begin{tabular}
      {l@{\extracolsep{\fill}}r@{\extracolsep{0pt}$\:\pm\:$}l@{}%
        @{\extracolsep{\fill}}r@{\extracolsep{0pt}$\,\pm\,$}l@{}%
        @{\extracolsep{\fill}}r@{\extracolsep{0pt}$\,\pm\,$}l@{}}
~~~~Source & \multicolumn{2}{c}{2 jets} & \multicolumn{2}{c}{3 jets}
                                        & \multicolumn{2}{c}{4 jets}
\vspace{0.03in} \\
\hline
$tb$+$tqb$ signal         &   139 &  18 &    63 &  10 &  21 &  5 \\
$W$+jets                  & 1,829 & 161 &   637 &  61 & 180 & 18 \\
$Z$+jets and dibosons     &   229 &  38 &    85 &  17 &  26 &  7 \\
${\ttbar}$                &   222 &  35 &   436 &  66 & 484 & 71 \\
Multijets                 &   196 &  50 &    73 &  17 &  30 &  6 \\ 
Total prediction~~        & 2,615 & 192 & 1,294 & 107 & 742 & 80 \\
Data                      & \multicolumn{2}{c}{2,579}
                          & \multicolumn{2}{c}{1,216}
			  & \multicolumn{2}{c}{724}      
\end{tabular}
\end{ruledtabular}
\end{table}


Systematic uncertainties arise from each correction factor or function
applied to the background and signal models. Most affect only the
normalization, but three corrections modify in addition the shapes of
the distributions; these are the jet energy scale corrections, the
tag-rate functions, and the reweighting of the distributions in
$W$+jets events. The largest uncertainties come from the jet energy
scale (the normalization part is (1.1--13.1)\% for signal and
(0.1--2.1)\% for background), the tag-rate functions (the
normalization part is (2.1--7.0)\% for single-tagged events and
(9.0--11.4)\% for double-tagged events), and the correction for
jet-flavor composition in $W$+jets events (13.7\%), with smaller
contributions from the integrated luminosity (6.1\%), jet energy
resolution (4.0\%), initial-state and final-state radiation
(0.6--12.6\%), $b$-jet fragmentation (2.0\%), {\ttbar} cross section
(12.7\%), and lepton efficiency corrections (2.5\%). All other
contributions have a smaller effect. The values given are the relative
uncertainties on the individual sources. The total uncertainty on the
background is (8--16)\% depending on the analysis channel.


After event selection, we expect single top quark events to constitute
(3--9)\% of the data sample. Since the uncertainty on the background
is larger than the expected signal, we improve discrimination by using
multivariate analysis techniques. We have developed three independent
analyses based on boosted decision trees (BDT)~\cite{decision-trees},
Bayesian neural networks (BNN)~\cite{bayesianNNs}, and the matrix
element (ME) method~\cite{matrix-elements}. Our application of these
techniques to D0's single top quark searches is described in
Refs.~\cite{d0-prl-2007} and \cite{d0-prd-2008}. The analyses
presented in this Letter differ from previous implementations in the
choice of input variables and some detailed tuning of each technique.


The BDT analysis has re-optimized the input
variables~\cite{variables-schwienhorst} into a common set of 64
variables for all analysis channels. The variables fall into five
categories, single-object kinematics, global event kinematics, jet
reconstruction, top quark reconstruction, and angular correlations.
Separate sets of trees are created with these variables for each
channel. The BNN analysis uses the RuleFitJF algorithm~\cite{rulefit}
to select the most sensitive of these variables, then combines 18--28
of them into a single separate discriminant for each channel. The ME
analysis uses only two-jet and three-jet events, divided into a
$W$+jets-dominated set and a {\ttbar}-dominated set. It includes
matrix elements for more background sources, adding {\ttbar}, $WW$,
$WZ$, and $ggg$ diagrams in the two-jet bin and $Wugg$ in the
three-jet bin, to improve background rejection.

\begin{figure}[!h!tbp]
\vspace{-0.05in}
\includegraphics[width=1.66in]{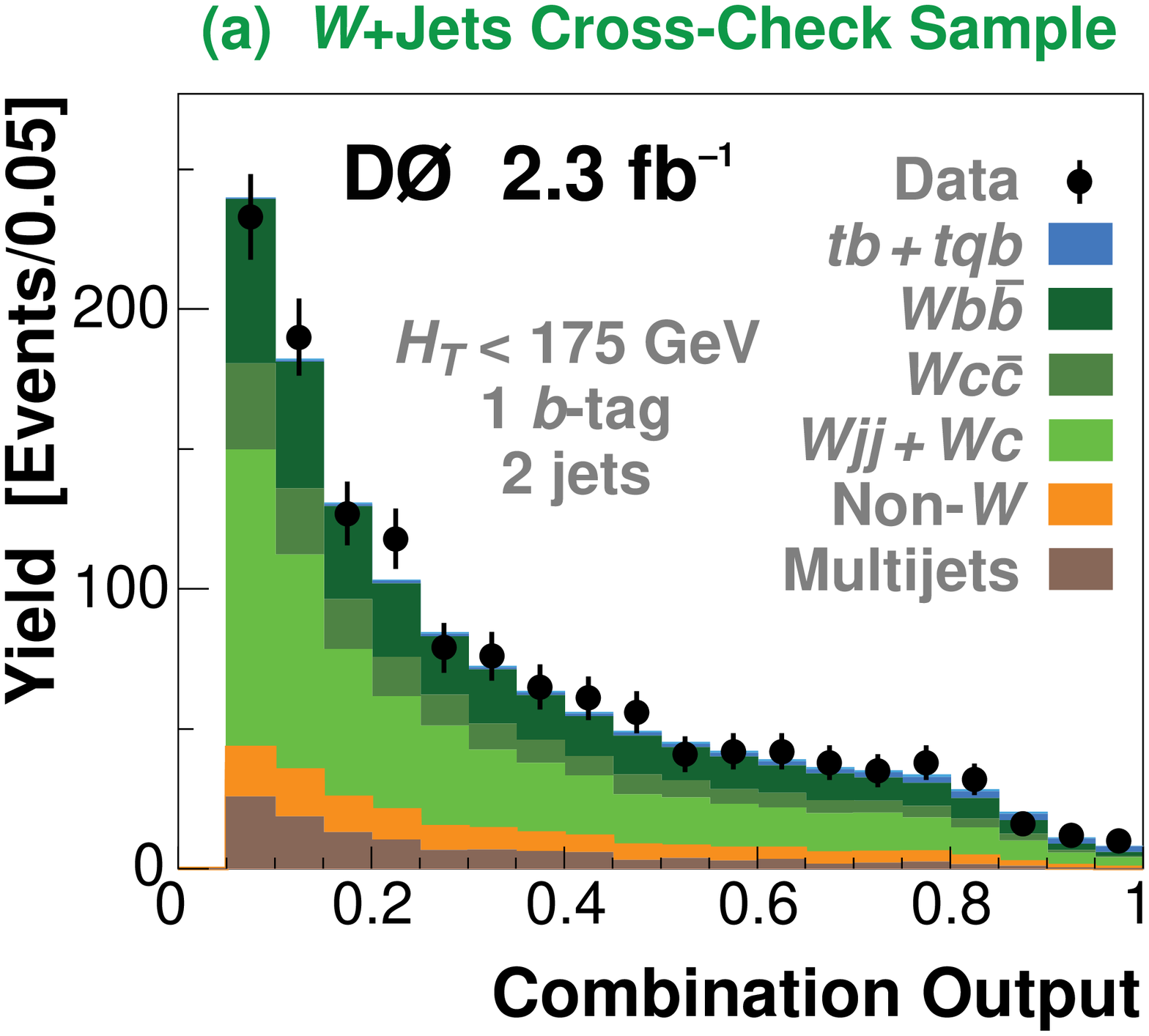}
\includegraphics[width=1.66in]{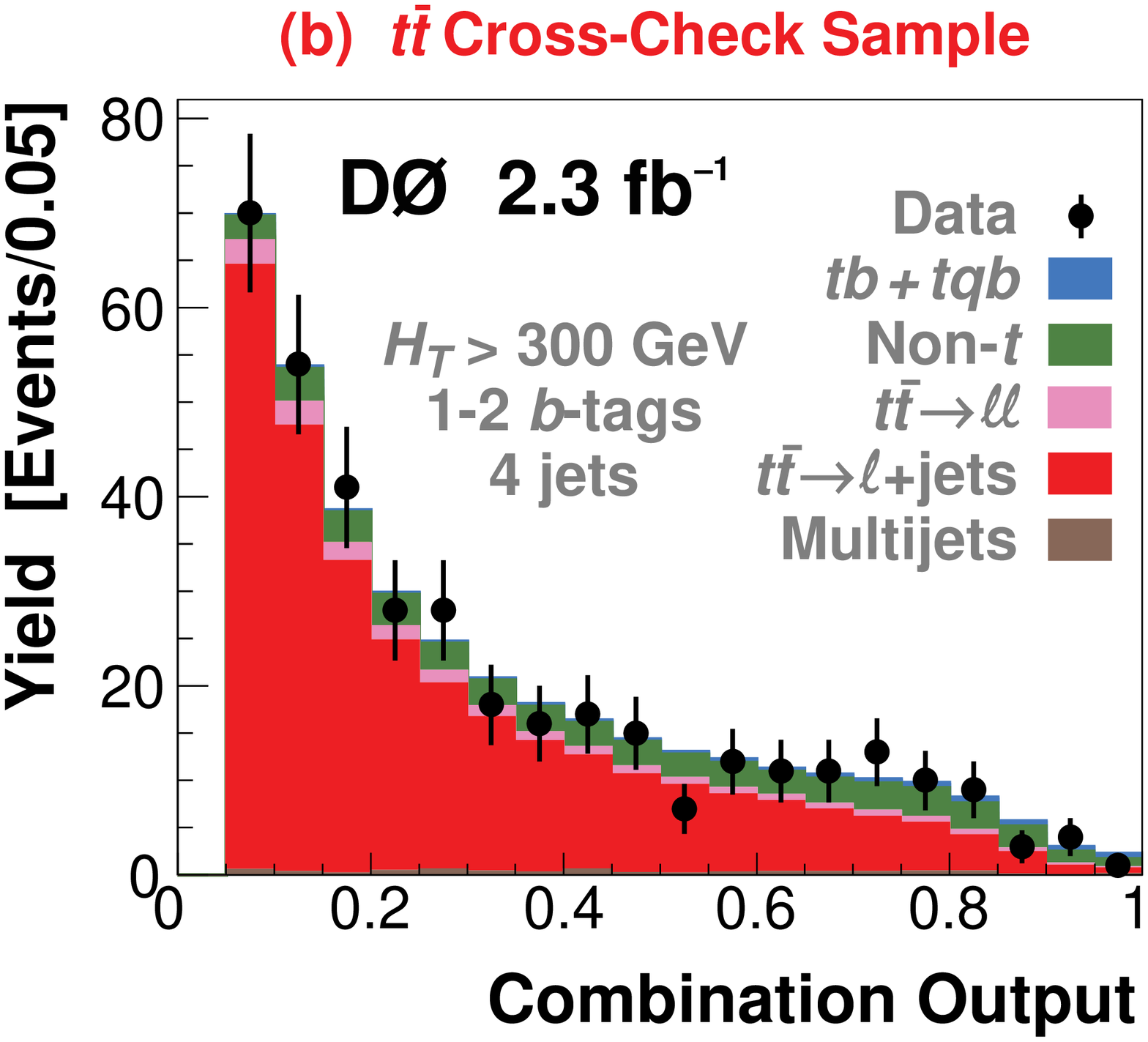}
\vspace{-0.1in}
\caption[crosschecks]{The combination discriminant outputs for
(a)~$W$+jets and (b)~{\ttbar} cross-check samples. $H_T$ is the scalar
sum of the transverse momenta of the final state objects (lepton, \met, 
and jets).}
\label{cross-checks}
\vspace{-0.1in}
\end{figure}

Each analysis uses the same data and background model and has the same
sources of systematic uncertainty. We test the analyses using
ensembles of pseudodatasets created from background and signal at
different cross sections to confirm linear behavior and thus an
unbiased cross section measurement. The analyses are also checked
extensively before $b$-tagging is applied, and using two control
regions of the data, one dominated by $W$+jets and the other by
{\ttbar} backgrounds, as shown in Fig.~\ref{cross-checks}. These
studies confirm that backgrounds are well modeled across the full
range of the discriminant output.

The cross section is determined using the same Bayesian approach as in
our previous studies~\cite{d0-prl-2007,d0-prd-2008}. This involves
forming a binned likelihood as a product over all bins and channels,
evaluated separately for each multivariate discriminant, with no cuts
applied to the outputs. The central value of the cross section is
defined by the position of the peak in the posterior density, and the
$68\%$ interval about the peak is taken as the uncertainty on the
measurement. Systematic uncertainties, including all correlations, are
reflected in this posterior interval.

We extract inclusive single top quark cross sections
$\sigma({\ppbar}{\rargap}tb+X,tqb+X)$ of $\sigma_{\rm BDT} =
3.74^{+0.95}_{-0.79}$~pb, $\sigma_{\rm BNN} =
4.70^{+1.18}_{-0.93}$~pb, and $\sigma_{\rm ME} =
4.30^{+0.99}_{-1.20}$~pb. The sensitivity of the analyses to a
contribution from single top quark production is estimated by
generating an ensemble of pseudodatasets that sample the background
model and its uncertainties, with no signal present. We measure a
cross section from each pseudodataset, and hence obtain the
probability that the SM cross section is reached. This provides
expected sensitivities (stated in terms of Gaussian standard
deviations, SD) of 4.3, 4.1, and 4.1~SD for the BDT, BNN, and ME
analyses respectively. The measured significances, obtained by
counting the number of pseudodatasets with cross sections at least as
large as the measured cross section, are 4.6, 5.2, and 4.9~SD
respectively.

\begin{figure}[!h!tbp]
\includegraphics[width=1.66in]{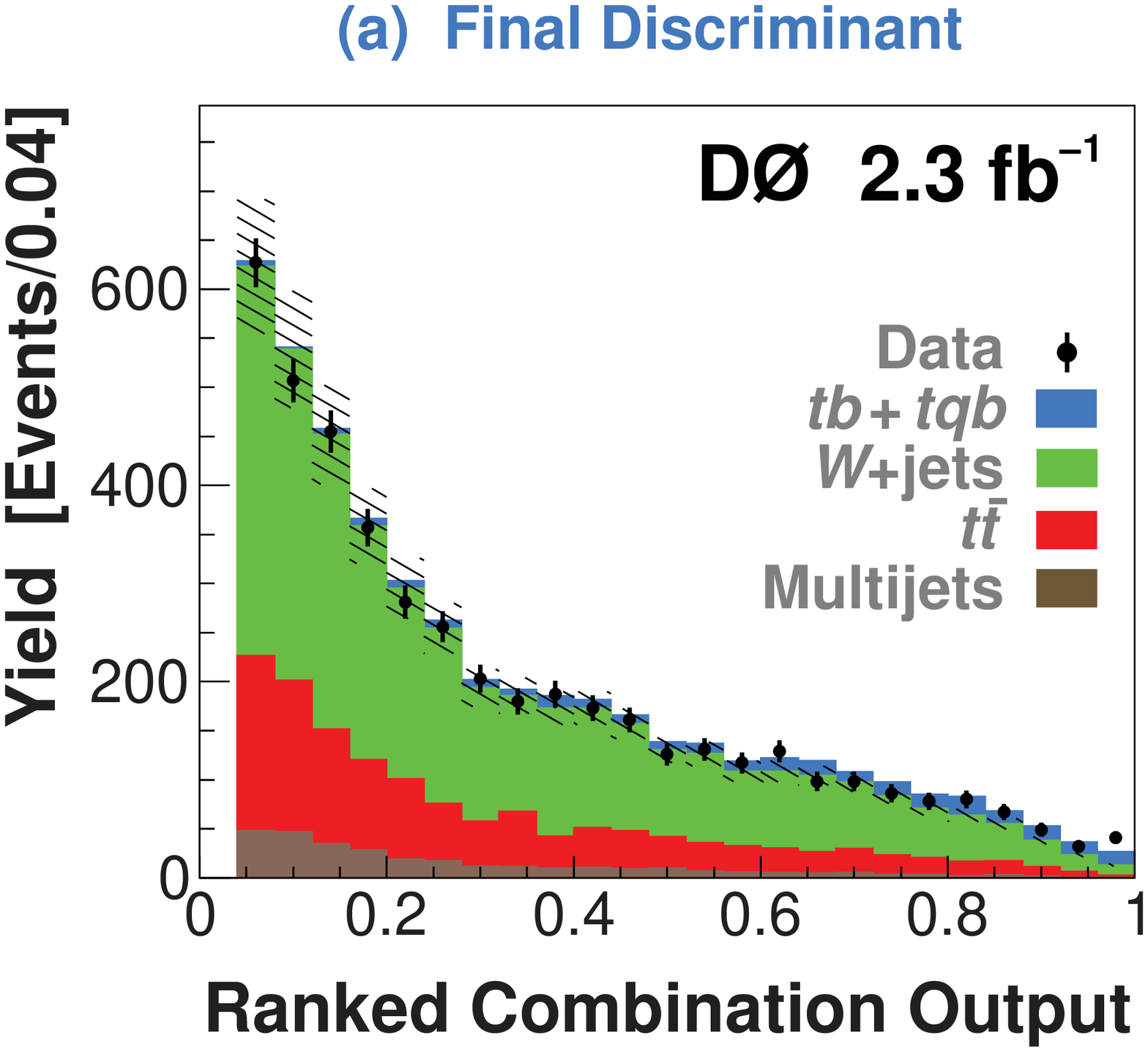}
\includegraphics[width=1.66in]{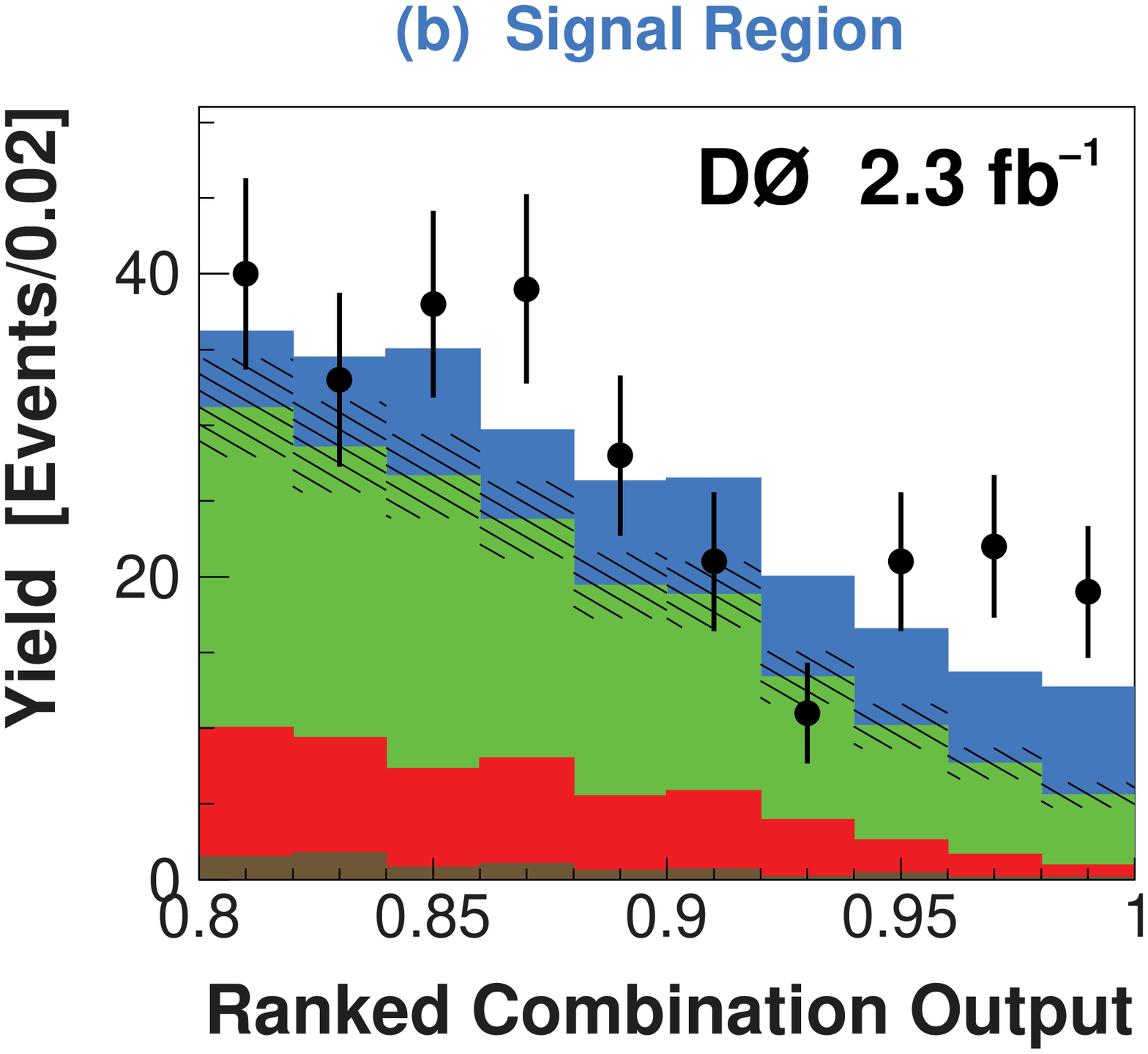}\\
\vspace{0.1in}
\includegraphics[width=1.66in]{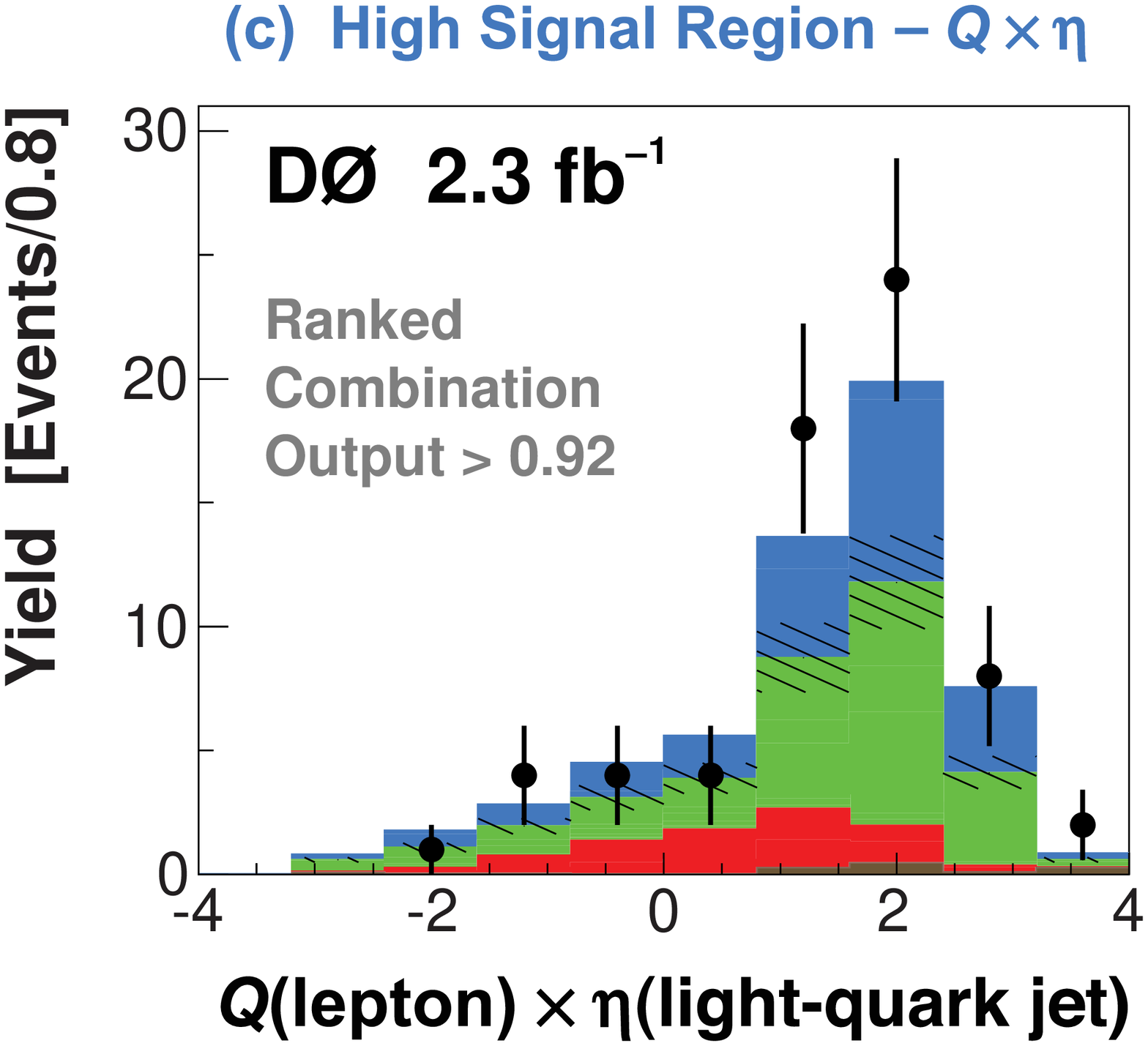}
\includegraphics[width=1.66in]{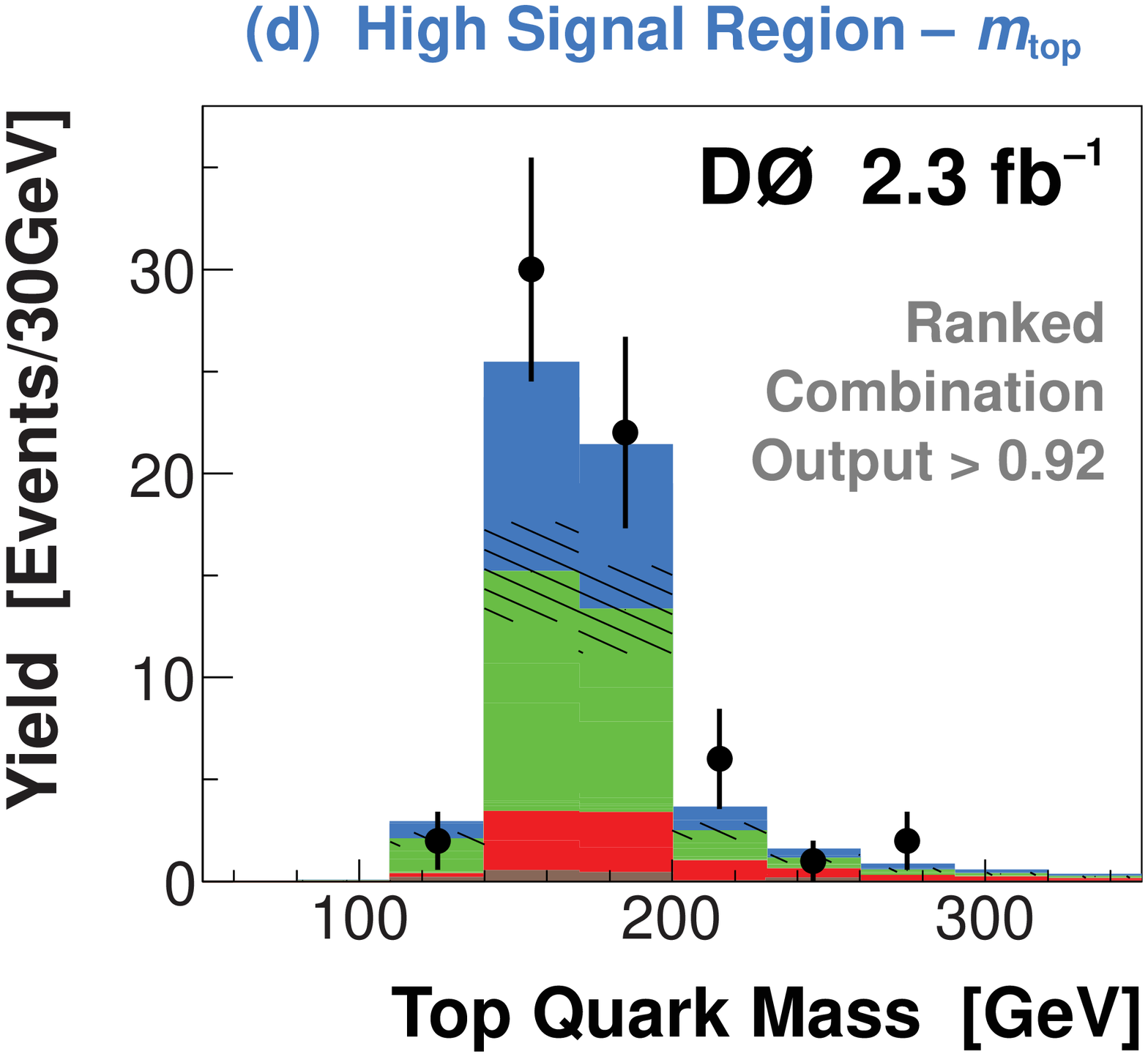}\\
\vspace{-0.1in}
\caption[discriminants]{Distribution of the combination output for
all 24 analysis channels combined, (a)~full range, and (b)~high signal
region. The bins have been ordered by their expected signal:background
ratio and the signal is normalized to the measured cross section. The
hatched band indicates the total uncertainty on the background. These
distributions are not used in the cross section measurement and are
for illustration only. For the ranked combination output $> 0.92$,
(c)~shows the distribution of lepton charge times pseudorapidity of
the leading not-$b$-tagged jet, and (d)~shows the reconstructed top
quark mass.}
\label{discriminant}
\end{figure}

The three multivariate techniques use the same data sample but are not
completely correlated: the correlation of the measured cross section
using pseudodatasets with background and SM signal is BDT:BNN = 74\%,
BDT:ME = 60\%, BNN:ME = 57\%. Their combination therefore leads to
increased sensitivity and a more precise measurement of the cross
section. We use the three discriminant outputs as inputs to a second
set of Bayesian neural networks, and obtain the combined cross section
and its signal significance from the new discriminant output. The
resulting expected significance is 4.5~SD. Figure~\ref{discriminant}
illustrates the importance of the signal when comparing data to
prediction.

The measured cross section is
\vspace{0.05in}

$\hspace{0.25in}
\sigma({\ppbar}{\rargap}tb+X,tqb+X) = 3.94 \pm 0.88 {\rm ~pb.}$

\begin{figure}[!h!tbp]
\includegraphics[width=1.66in]{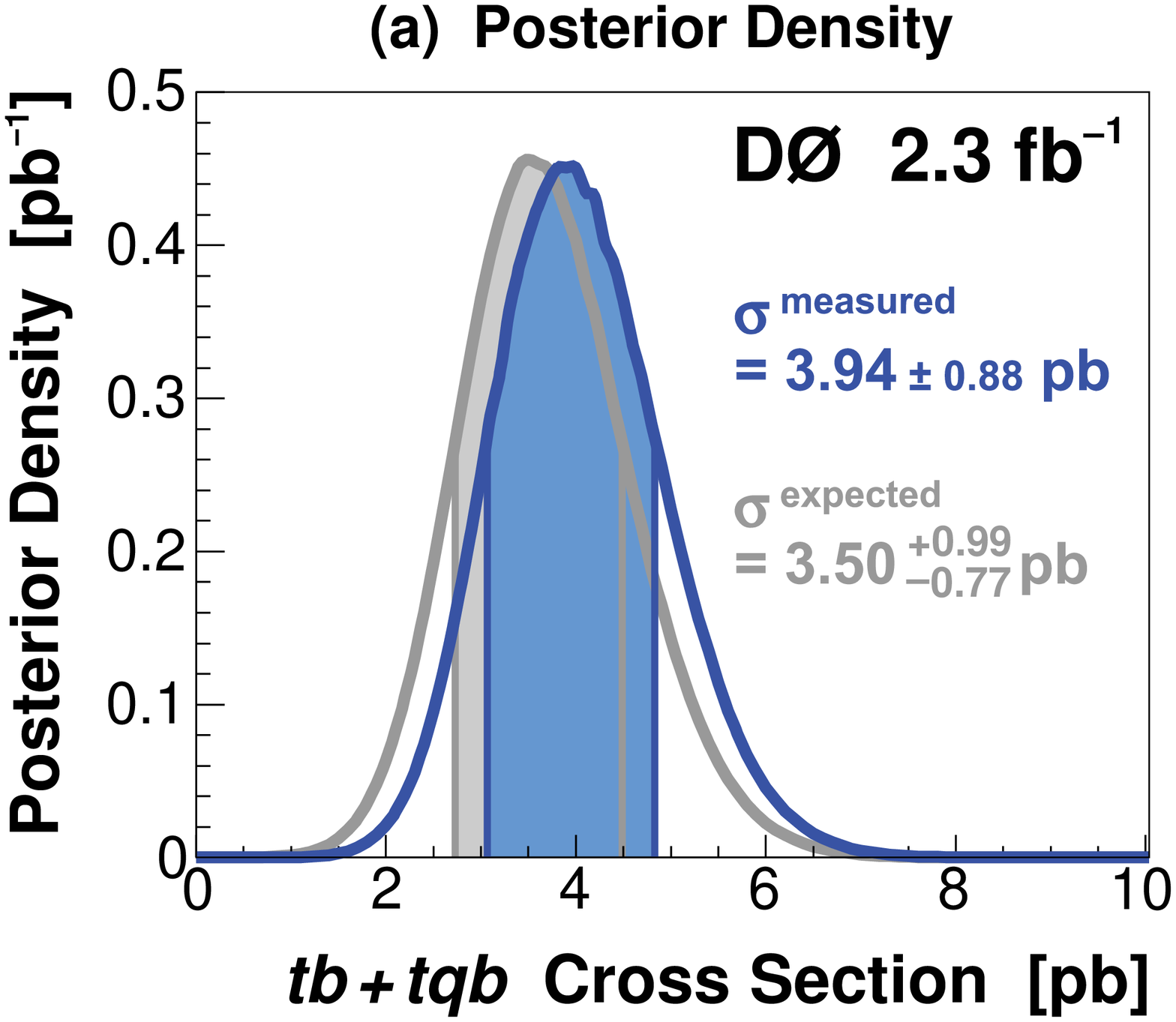}
\includegraphics[width=1.66in]{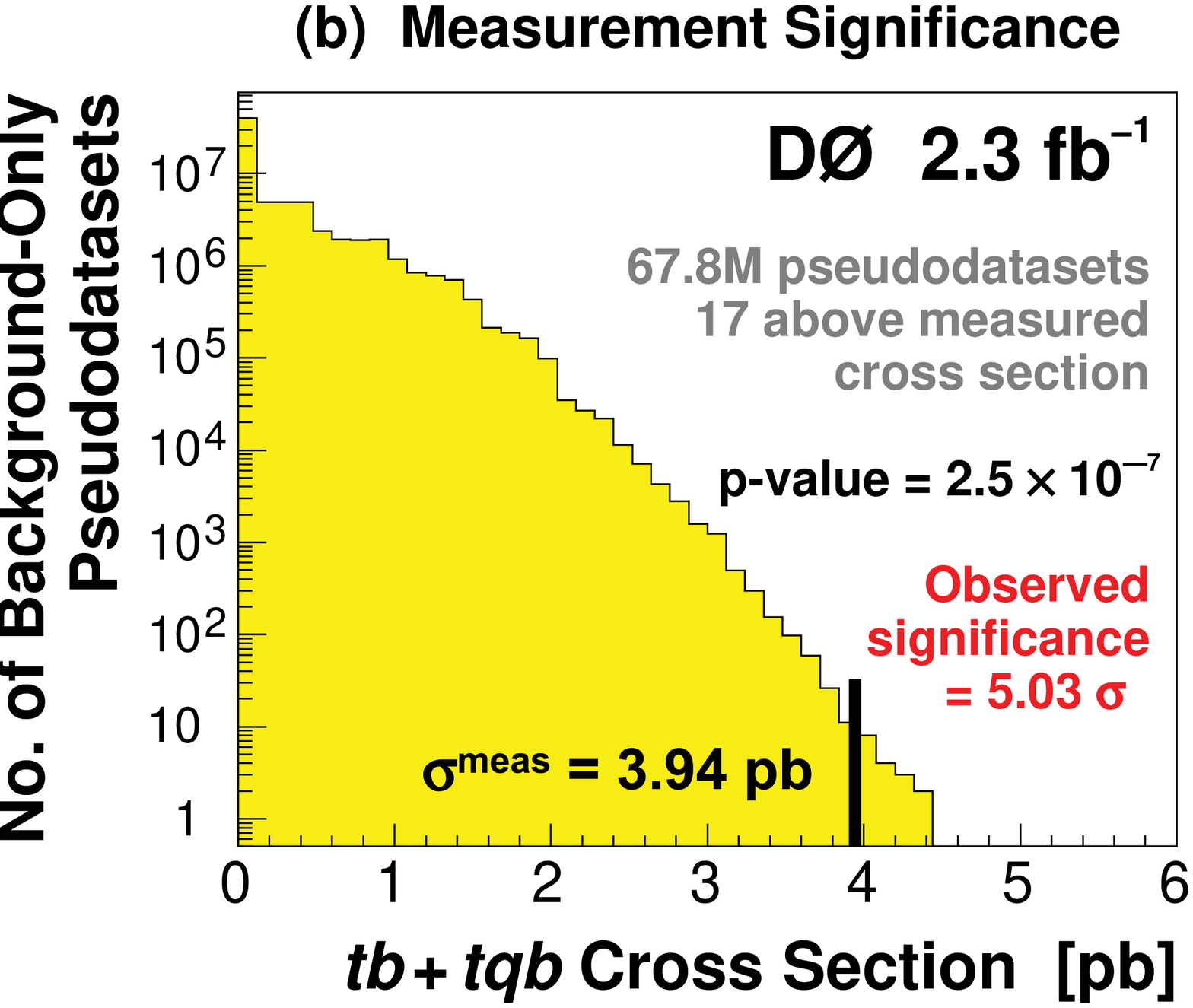}
\vspace{-0.1in}
\caption[posterior]{(a) Expected SM and measured Bayesian posterior
probability densities for the $tb$+$tqb$ cross section. The shaded
regions indicate one standard deviation above and below the peak
positions. (b) Measured cross sections using the ensemble of
background-only pseudodatasets (containing full systematics and no
signal) used to measured the significance of the result.}
\label{posteriors-significance}
\vspace{-0.1in}
\end{figure}

\vspace{0.05in}
\noindent The measurement has a $p$-value of $2.5 \times 10^{-7}$,
corresponding to a significance of 5.0~SD. The expected and measured
posterior densities and the background-only pseudodataset measurements
are shown in Fig.~\ref{posteriors-significance}.

We use the cross section measurement to determine the Bayesian
posterior for $|V_{tb}|^2$ in the interval [0,1] and extract a limit
of $|V_{tb}| > 0.78$ at 95\% C.L. within the
SM~\cite{d0-prd-2008}. When the upper constraint is removed, we
measure $|V_{tb}f_1^L| = 1.07 \pm 0.12$, where $f_1^L$ is the
strength of the left-handed $Wtb$ coupling.


In summary, we have measured the single top quark production cross
section using 2.3~fb$^{-1}$ of data at the D0 experiment. We measure a
cross section for the combined $tb$+$tqb$ channels of $3.94 \pm
0.88$~pb. Our result provides an improved direct measurement of the
amplitude of the CKM matrix element $V_{tb}$. The measured single top
quark signal corresponds to an excess over the predicted background
with a significance of 5.0~SD --- observation of
single top quark production.


%
We thank the staffs at Fermilab and collaborating institutions, 
and acknowledge support from the 
DOE and NSF (USA);
CEA and CNRS/IN2P3 (France);
FASI, Rosatom and RFBR (Russia);
CNPq, FAPERJ, FAPESP and FUNDUNESP (Brazil);
DAE and DST (India);
Colciencias (Colombia);
CONACyT (Mexico);
KRF and KOSEF (Korea);
CONICET and UBACyT (Argentina);
FOM (The Netherlands);
STFC (United Kingdom);
MSMT and GACR (Czech Republic);
CRC Program, CFI, NSERC and WestGrid Project (Canada);
BMBF and DFG (Germany);
SFI (Ireland);
The Swedish Research Council (Sweden);
CAS and CNSF (China);
and the
Alexander von Humboldt Foundation (Germany).

\vspace{-0.3in}



\end{document}